\documentstyle[aps,floats,psfig]{revtex}

\begin{document}
\psfigurepath{.:plot:figure}
\twocolumn[\hsize\textwidth\columnwidth\hsize\csname @twocolumnfalse\endcsname
\preprint{draft rsg8.tex, \today}
\title{\bf Unconventional ferromagnetic and spin-glass states of the
reentrant spin glass Fe$_{0.7}$Al$_{0.3}$}
\author{Wei Bao,$^{1,2}$\cite{byline} S. Raymond,$^{1,3}$ S. M. Shapiro,$^{1}$ 
K. Motoya,$^{4}$ B. F\aa k, $^{3}$ and R. W. Erwin$^{5}$}
\address{$^{1}$Brookhaven National Laboratory, Upton, NY 11973\\
$^{2}$Los Alamos National Laboratory, Los Alamos, NM 87545\\
$^{3}$CEA-Grenoble, DRFMC-SPSMS, 38054 Grenoble Cedex 9, France\\
$^{4}$Science University of Tokyo, Noda 278, Japan\\
$^{5}$National Institute of Standards and Technology, Gaithersburg, MD 20899
}
\date{\today}
\maketitle
\begin{abstract}
Spin excitations in a single crystal of Fe$_{0.7}$Al$_{0.3}$
were investigated over a wide range of energy and reciprocal space 
with inelastic neutron scattering.
In the ferromagnetic phase, propagating spin wave modes 
become paramagnon-like
diffusive modes beyond a critical wave vector {\bf q}$_0$, 
indicating substantial disorder in the long-range ordered state. 
In the spin glass phase, the spin dynamics are strongly
{\bf q}-dependent, suggesting
remnant short-range spin correlations. 
\end{abstract}
\vskip2pc]


\narrowtext

One class of disordered ferromagnets show reentrant spin-glass
behavior\cite{sg_jam}: magnetization measurements
suggest that the materials change
from paramagnetic to ferromagnetic (FM) at the Curie temperature, T$_C$.
Upon further lowering the temperature, the spins are progressively
frozen below a freezing temperature, T$_f$. 
The low temperature spin frozen state is called 
a reentrant spin-glass (SG) or mixed state, 
in which ferromagnetic order is argued to coexist with 
spin-glass order\cite{sgt_gt}.
For comparison, in a diluted spin-glass, spins freeze directly from
a paramagnetic state.
Spin waves, the expected collective excitation modes from ferromagnetic 
long range order, 
have been studied only at small wave vector, {\bf Q},
using inelastic neutron scattering
in the FM phase of the reentrant spin-glasses 
FeCr, AuFe, FeAl, NiMn, Fe(Ni,Mn) alloys and amorphous 
(Fe,{\it T})$_{75}$P$_{16}$B$_5$Al$_3$ ({\it T}=Mn, Ni, or 
Cr)\cite{rsg_sfpp,rsg_murani,rsg_motoya,rsg_asbc,rsg_lms}.
The spin waves become broad, decrease in energy,
and a quasielastic component appears
when the temperature approaches T$_f$.
This short-range quasielastic component continues to increase with
lowering temperature in the SG state, as observed in diluted 
spin-glass such as Cu{\it Mn}\cite{sg_muranit}.

Spin-glasses are generally associated with large degeneracy
of the magnetic states
caused by disorder, frustration\cite{rev_part} or both.
In a reentrant spin-glass, the fact that the FM state occurs 
at higher temperatures than the SG state 
suggests a larger entropy for the FM state than 
for the SG state\cite{sg_jam}. 
This apparent, counter-intuitive situation remains a major 
mystery in the field of disordered magnetic
systems\cite{sg_jam}.
In this work, using a single crystal, we explore
spin dynamics in a greatly expanded
{\bf Q} and $\omega$ range beyond previous neutron 
scattering studies\cite{rsg_sms}.
We find that the FM state of the reentrant spin-glass 
Fe$_{0.7}$Al$_{0.3}$ has a qualitatively different spin dynamics
behavior
from that of a conventional ferromagnetic state. It consists of
a mixture of low energy propagating
spin waves at small wave vectors and diffusive paramagnon-like spin
fluctuations at large wave vectors. This is strikingly reminiscent
of the phonon behavior in structural glasses\cite{gls_si}.
This confirms that although there is ferromagnetic order,
substantial disorders exist at finite length scales in this
anomalous ferromagnetic state. 
In addition, a strongly {\bf Q}-dependent spin excitation spectrum 
exists in the SG state, 
suggesting remnant short-range spin correlations. 
Preliminary results were reported at a conference\cite{stephane}.

The single crystal sample of Fe$_{0.7}$Al$_{0.3}$ 
(Fe$_{3-\delta}$Al$_{1+\delta}$, $\delta=0.2$) used in this 
study has a volume of $\sim$1 $cm^3$ with a mosaic $\sim$$1.5^o$. 
It has the face-centered-cubic (fcc) DO$_3$ structure
(space group $Fm\overline{3}m$, No.\ 225) with four crystallographic sites
(refer to Fig.~\ref{fig_str}): 
\begin{figure}[bt]
\centerline{
\psfig{file=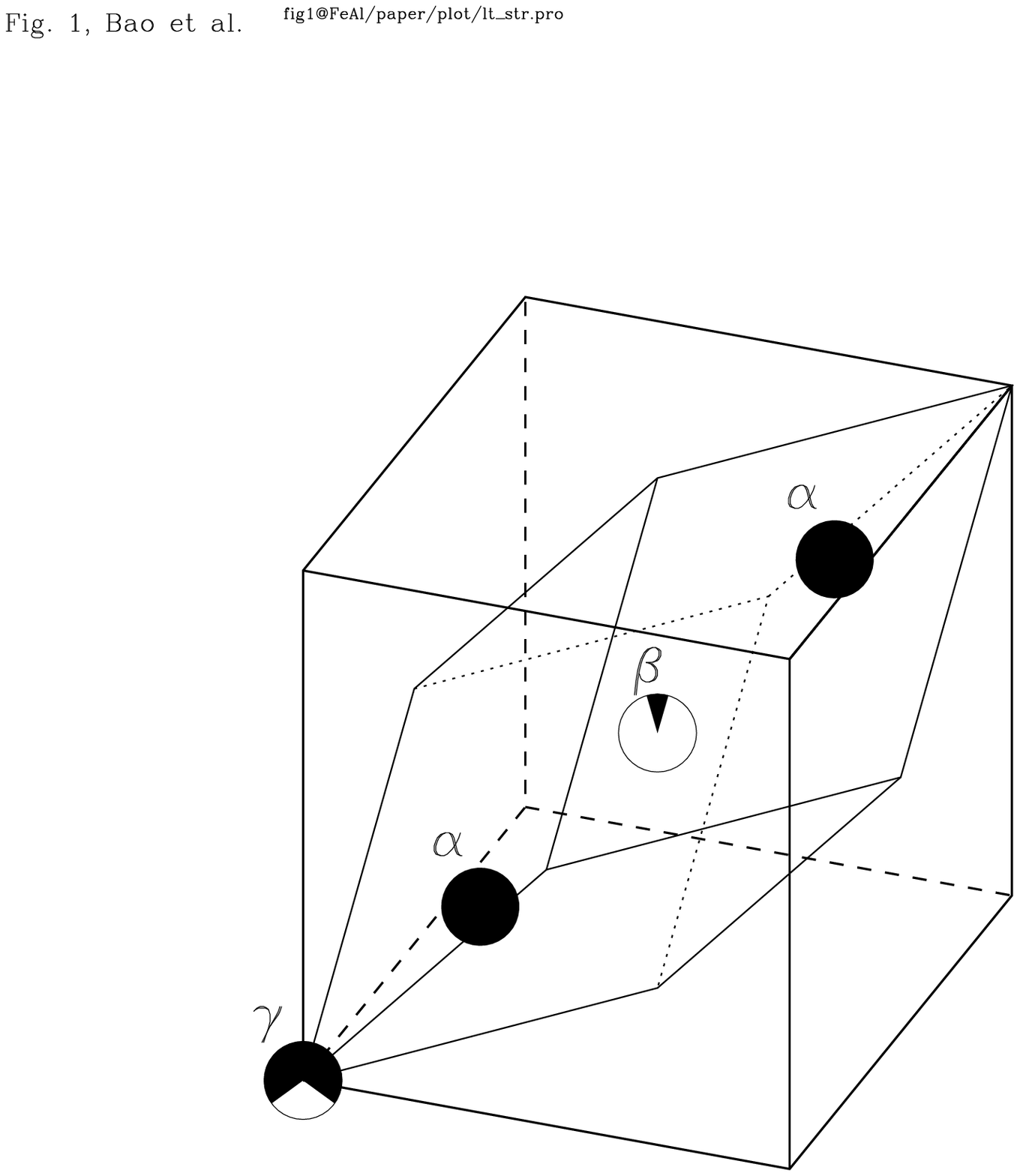,width=.45\columnwidth,angle=0,clip=}}
\caption{The fcc lattice structure of Fe$_{0.7}$Al$_{0.3}$.
Only atoms inside the primitive unit cell are shown.
The $\alpha$ sites are occupied by Fe (black) only.
The $\beta$ site is predominately occupied by Al (white)
while the $\gamma$ site is mostly occupied by Fe.
}
\label{fig_str}
\end{figure}
the $\alpha$ sites 
($\slantfrac{1}{4}$,$\slantfrac{1}{4}$,$\slantfrac{1}{4}$) 
and ($\slantfrac{3}{4}$,$\slantfrac{3}{4}$,$\slantfrac{3}{4}$)
are occupied only by Fe, the $\beta$ site 
($\slantfrac{1}{2}$,$\slantfrac{1}{2}$,$\slantfrac{1}{2}$)
is occupied predominately by
Al with minority Fe ($\sim$9\%), and the $\gamma$ site (000)
is occupied 
predominately by Fe ($\sim$70\%) with minority Al\cite{rsg_cable}.
The lattice parameter is $a=5.803\AA$ at the room temperature,
and there are four formula units in each unit cell.
The Curie temperature is T$_C \approx 520$~K and 
the spin freezing temperature
T$_f \approx 80$~K\cite{rsg_shull}. 

Neutron scattering experiments were performed at the cold neutron
triple-axis spectrometers H9A at the High Flux Beam
Reactor (HFBR) of BNL, IN12 at ILL Grenoble, thermal neutron triple-axis
spectrometers H7 of HFBR, and BT2 at NIST.
Pyrolytic graphite (PG) was used as a monochromator and 
analyzer. A cold beryllium filter was used for cold neutron
measurements and a PG filter was used for thermal neutron 
measurements to reduce higher order neutrons. Spectrometer 
configurations used in experiments are specified in the figures.
The sample was placed 
inside an aluminum can filled with He exchange gas 
in a cryostat. It was aligned with the ($hhl$) zone 
coinciding with the scattering plane. 
We focus in this paper on results from
scans along the [111] direction near Bragg points
(111) and (000).

The intensity of the magnetic neutron scattering was measured against
a flux monitor placed between the sample and the exit collimator 
for monochromator. It can be expressed\cite{bibnjc} as
\begin{equation}
I({\bf Q},\omega)=A\cdot f(k_i) \cdot \frac{k_f^3}{\tan \theta_A}
\cdot |F({\bf Q})|^2 \cdot \overline{S}({\bf Q},\omega),
\end{equation}
where A is approximately a constant for a given spectrometer
configuration, $k_i$ and $k_f$ are, respectively, the initial
and final wave number for neutrons, 
$2\theta_A$ is the scattering
angle for analyzer, $f(k_i)$ is a correction factor
for high order neutrons registered at the flux monitor
in a fixed $k_f$ configuration,
$|F({\bf Q})|^2$ is atomic form factor for Fe,
and $\overline{S}({\bf Q},\omega)$ is the convolution of the
dynamic spin correlation function $S({\bf Q},\omega)$ with 
spectrometer resolution function.
The polarization factor for ${\bf Q}$ along [111]
has been absorbed in $A$.
By measuring neutron scattering intensity as a function of momentum 
transfer $\hbar {\bf Q}$ and energy transfer $\hbar \omega$,
$\overline{S}({\bf Q},\omega)$ can be directly determined. 
For the configurations we used,  the four-dimensional
convolution yields $\overline{S}({\bf Q},\omega)
\approx S({\bf Q},\omega)$ for $|\hbar \omega|$ greater than 
the energy resolution\cite{note}.
Factoring out the thermal occupation factor, 
the imaginary part of the
generalized dynamic magnetic susceptibility is given by
\begin{equation}
\chi''({\bf Q},\omega)=\pi \left( 1-e^{-\hbar\omega/k_B T}\right)
S({\bf Q},\omega). \label{eq_fdt}
\end{equation}
The neutron scattering data shown in this paper are normalized to 
yield either the $S({\bf Q},\omega)$ or $\chi''({\bf Q},\omega)$.

\begin{figure}[bt]
\centerline{
\psfig{file=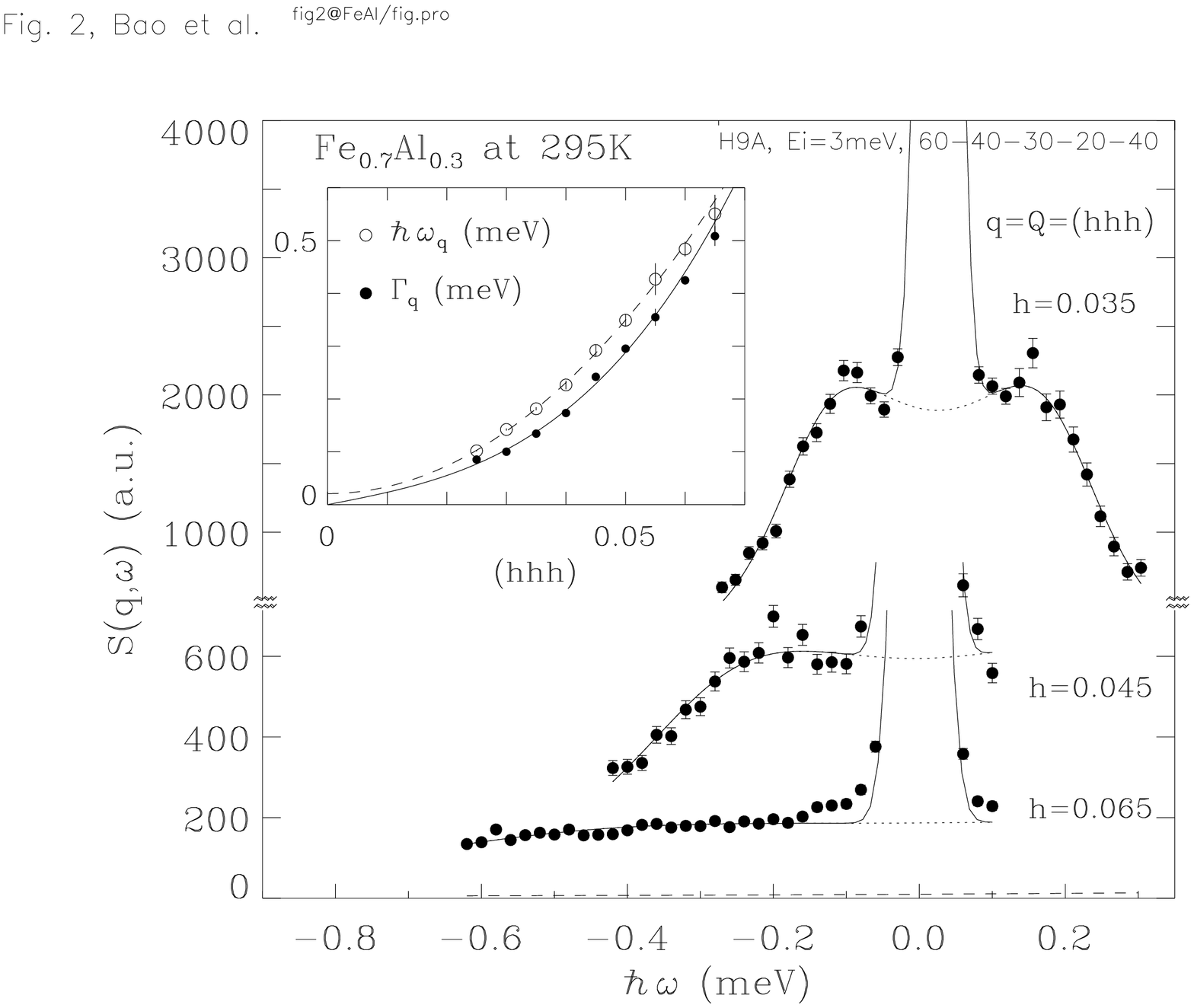,width=\columnwidth,angle=90,clip=}}
\caption{Small angle inelastic energy scans measured at several {\bf Q} 
values in
the FM phase of Fe$_{0.7}$Al$_{0.3}$.
Insert: spin wave energy $\hbar\omega_{\bf q}$ and 
damping energy $\Gamma_{\bf q}$ 
[Eq.~(\ref{eq_dho})]
vs {\bf q}=(hhh).
}
\label{fig_fwd}
\end{figure}

We present first our results at 295~K well within the FM phase.  
Fig.~\ref{fig_fwd} shows a few examples of
constant ${\bf Q}=(hhh)$ scans near (000). 
The background (refer to the dashed line) is negligible compared to
signal. The resolution limited elastic peak at $\hbar\omega=0$
has been studied before\cite{rsg_motoya}
and it will not be discussed here.
As observed in previous works near the forward 
direction\cite{rsg_sfpp,rsg_murani,rsg_motoya,rsg_asbc,rsg_lms}, 
$S({\bf Q},\omega)$
peaks at a finite energy. With increasing {\bf Q}, the peak energy
increases while the damping of the peak becomes stronger. 
For $h>0.06$, the spectra are overdamped.

We model the magnetic excitations with a form which is equivalent to
the damped harmonic oscillator\cite{dho}:
\begin{equation}
\chi''({\bf q},\omega)= C
\left(
\frac{\Gamma_{\bf q}}{\hbar^2(\omega-\omega_{\bf q})^2+\Gamma_{\bf q}^2}
-\frac{\Gamma_{\bf q}}{\hbar^2(\omega+\omega_{\bf q})^2+\Gamma_{\bf q}^2}
\right)   \label{eq_dho}
\end{equation}
where $C$ is a constant oscillator strength, and 
we have chosen ${\bf Q}={\bbox \tau} +{\bf q}$, where ${\bbox \tau}$
is the (111) or (000) Bragg point, so that ${\bf q}$ is defined 
within a Brillouin zone.
From a standard least-squares fit to each scan, 
the energy of the spin wave mode, $\hbar \omega_{\bf q}$, and the damping 
energy, $\Gamma_{\bf q}$, are obtained
(refer to the insert in Fig.~\ref{fig_fwd}). 
The $\hbar \omega_{\bf q}$ can be described by
\begin{equation}
\hbar \omega_{\bf q}= D q^2+\Delta \label{eq_mode}
\end{equation}
with the stiffness constant $D=37.4(3)$ meV$\AA^2$ 
and $\Delta=0.021(2)$ meV (refer to the dashed line
in the insert in Fig.~\ref{fig_fwd}). 
For comparison, $D=101$ meV$\AA^2$ for stoichiometric Fe$_3$Al.
The value for $\Delta$ is much
smaller than the energy resolution ($\sim$0.07~meV)
so it may not be significantly
different from $\Delta=0$. Even in this propagating
spin wave regime,
the damping energy, $\Gamma_{\bf q}$, 
is close to $\hbar \omega_{\bf q}$ in magnitude.

\begin{figure}[bt]
\centerline{
\psfig{file=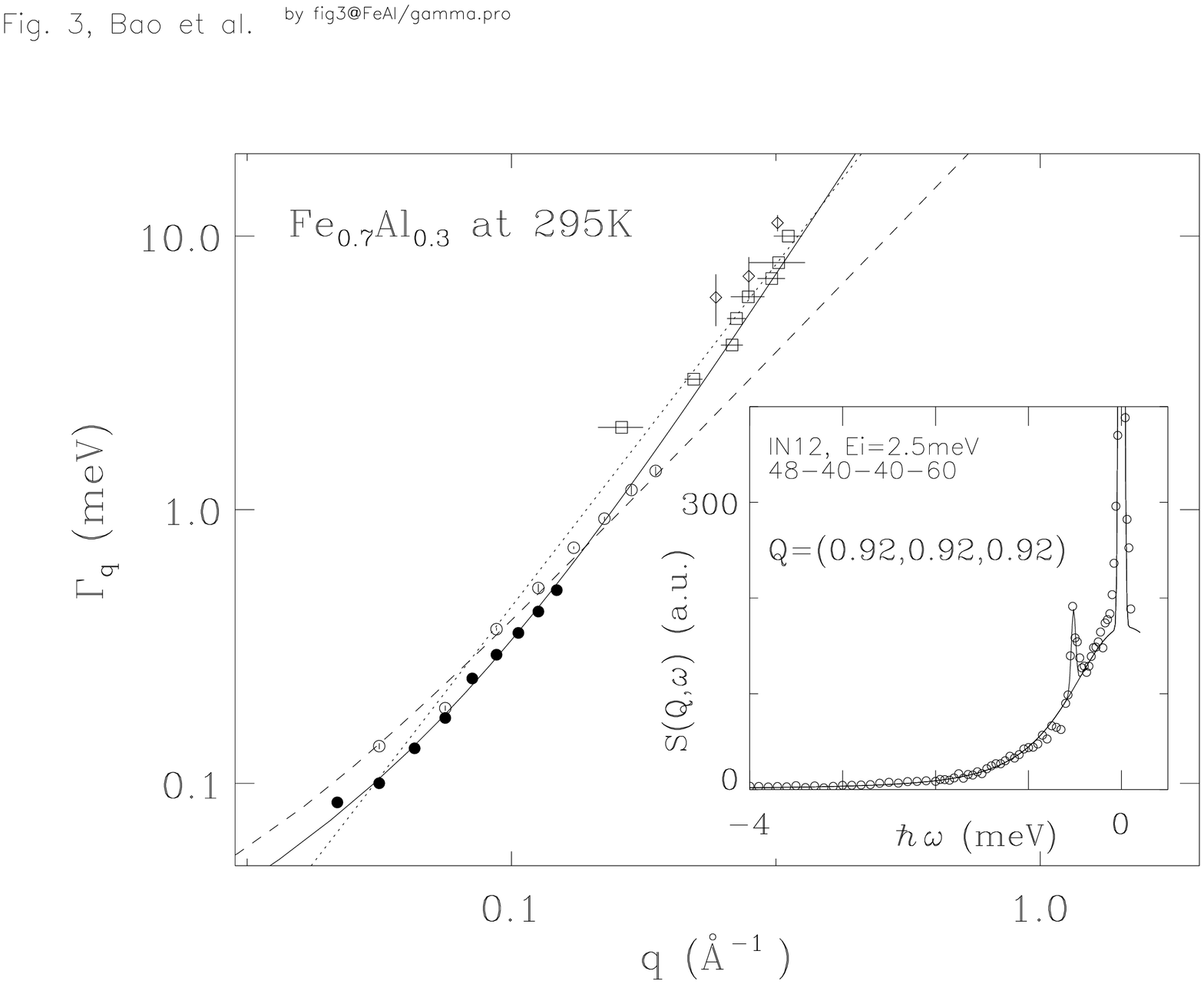,width=\columnwidth,angle=90,clip=}}
\caption{Damping energy $\Gamma_{\bf q}$ vs q. 
The solid line is a fit to Eq.~(\ref{eq_damp}) and the dotted line is
$\Gamma_{\bf q} \propto q^{2.5}$.
The solid circles are from measurements near (000)
at H9A with fixed $E_i=3$~meV and horizontal collimations
60$^\prime$-40$^\prime$-30$^\prime$-20$^\prime$-40$^\prime$; 
the open circles near (111) at IN12 with
$E_i=2.5$~meV and 48$^\prime$-40$^\prime$-40$^\prime$-60$^\prime$; 
the squares near (111) at H7 with $E_f=14.7$~meV and 
40$^\prime$-20$^\prime$-40$^\prime$-40$^\prime$; and the diamonds
near (111) at BT2 with $E_f=14.7$~meV and 
60$^\prime$-20$^\prime$-20$^\prime$-80$^\prime$.
The dashed line is $\hbar\omega_{\bf q}$ [Eq.~(\ref{eq_mode})]
for reference.
Insert: a const-{\bf Q} scan [{\bf q}=(0.08,0.08,0.08)]
near (111) measured in the FM phase.
}
\label{fig_gamma}
\end{figure}

The range of energy scans is
limited by the energy and momentum conservation conditions 
\[-\frac{\hbar^2}{2m}(2k_i\, Q+Q^2) <\hbar
\omega< \frac{\hbar^2}{2m}(2k_i\, Q-Q^2).\] 
For {\bf Q} larger than (0.06,0.06,0.06) (refer to  Fig.~\ref{fig_fwd}), 
the accessible energy range can hardly cover $\hbar \omega_{\bf q}$ or 
$\Gamma_{\bf q}$. To obtain reliable measurement of
$S({\bf q},\omega)$, we measure near the Bragg point
(111) where a much more extended energy range can be achieved 
with comparable energy resolution. An example for 
${\bf q}=(0.08,0.08,0.08)$ is shown in
the insert in Fig.~\ref{fig_gamma}. 
The sharp peak at $-0.52$~meV does not appear in a similar
scan near (000), thus it is not part of the magnetic spectrum we
are studying. Its origin is still under investigation.
The broad magnetic excitations is now clearly
overdamped.

For such overdamped magnetic excitations at 
large $q$, the value for
$\hbar \omega_{\bf q}$ cannot be obtained reliably.
We fixed $\hbar \omega_{\bf q}$ according to (\ref{eq_mode}) 
to estimate $\Gamma_{\bf q}$. The 
fitting was not very sensitive to $\hbar \omega_{\bf q}$.
Values of $\Gamma_{\bf q}$
from this and other configurations, spanning more than
two orders of magnitude in energy, 
are shown in the main part of Fig.~\ref{fig_gamma} 
using a log-log scale. Roughly, $\Gamma_{\bf q}\propto q^{2.5}$ 
(the dotted line).  However, the {\bf q} dependence of $\Gamma_{\bf q}$ 
may be better described by
\begin{equation}
\Gamma_{\bf q}=\gamma_{F}\, q^3 \left[ 1+ \left(\frac{\kappa}{q}
\right)^2 \right] \label{eq_damp} 
\end{equation}
with $\gamma_{F}=218(21)$ meV$\AA^3$ and an inverse length scale
$\kappa=0.073(2) \AA^{-1}$ (refer to the solid line). 
This {\bf q} dependence for damping is very similar to that
for paramagnetic spin fluctuations as observed at $T>T_C$
in ferromagnet MnSi, Pd$_2$MnSn, Ni 
and Fe\cite{MnSi_ishc}.
The damping in the {\em ferro}magnetic state of Fe$_{0.7}$Al$_{0.3}$,
thus, appears to share the same characteristics with damping in 
the thermally
disordered {\em para}magnetic state of ferromagnets.

\begin{figure}[bt]
\centerline{
\psfig{file=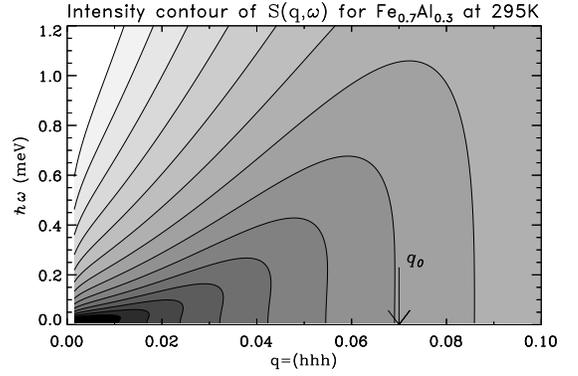,width=.9\columnwidth,angle=90,clip=}}
\caption{Intensity contour of $S({\bf q},\omega)$ in units of $C$meV
[refer to Eq.~(\ref{eq_dho})] using the experimentally determined 
$\omega_{\bf q}$ and $\Gamma_{\bf q}$ measured at 295~K. Each 
fading contour
level designates a decrease of one order of magnitude. The arrow
indicates the critical wave vector.}
\label{fig_map}
\end{figure}

\begin{figure}[bt]
\centerline{
\psfig{file=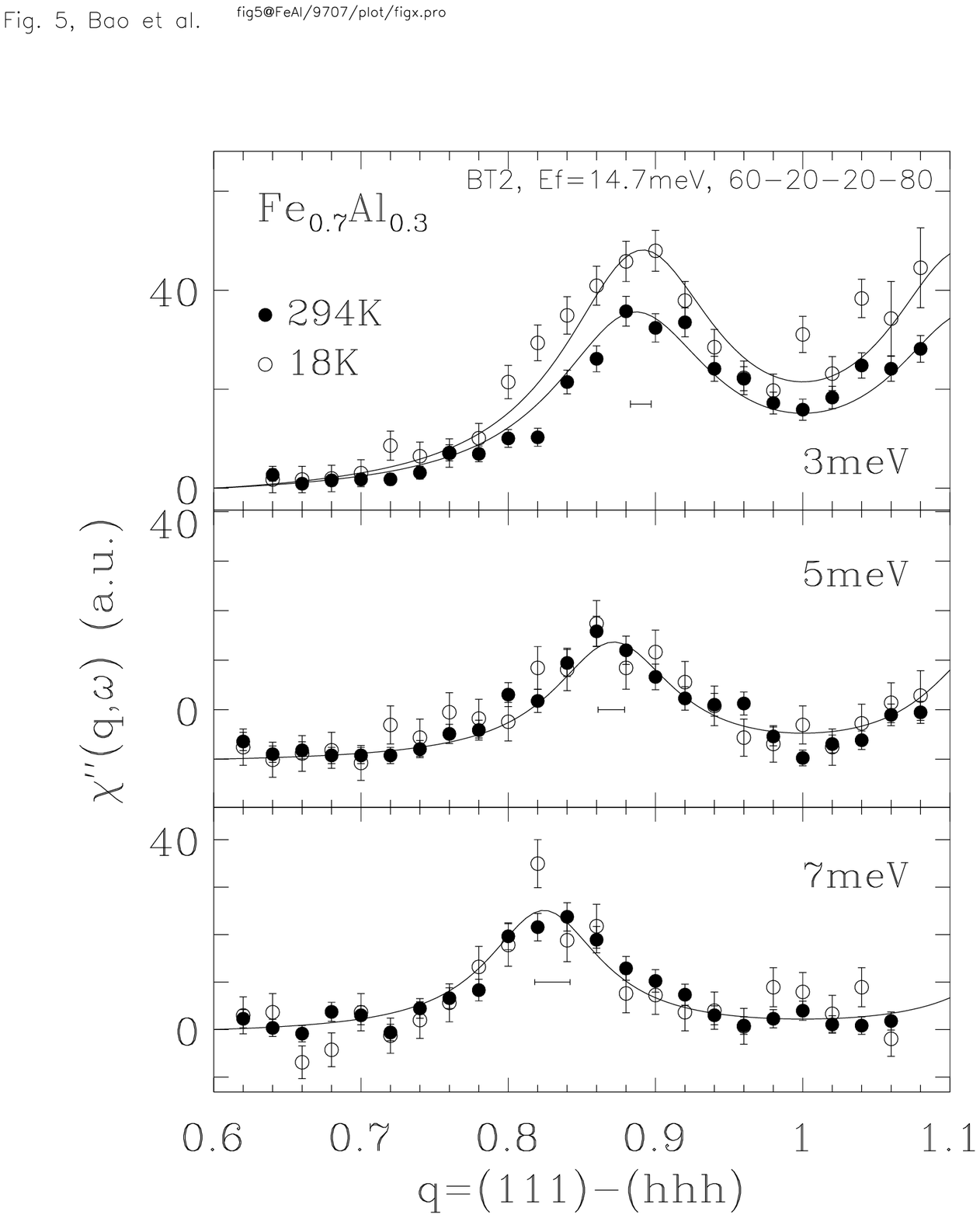,width=.7\columnwidth,angle=0,clip=}}
\caption{$\chi''({\bf q},\omega)$ determined from constant E scans
for $\hbar \omega=3$, 5 and 7~meV. 
The solid circles were measured at 294~K in the FM
phase, and the open circles at 18~K in the SG phase.
The horizontal bars indicate the full width at half maximum of the
projected instrument resolution function.}
\label{fig_cq}
\end{figure}

To better appreciate this unusual magnetic excitation spectrum,
intensity contours of $S({\bf q},\omega)$ for Fe$_{0.7}$Al$_{0.3}$
at 295~K, using Eq.~(\ref{eq_fdt}),
(\ref{eq_dho}) and experimentally
determined $\omega_{\bf q}$
and $\Gamma_{\bf q}$ in (\ref{eq_mode}) and (\ref{eq_damp}),
are shown in Fig.~\ref{fig_map}. There exists a critical wave number, $q_0$,
where $\hbar\omega_{{\bf q}_0} = \Gamma_{{\bf q}_0}$. 
At room temperature,
\begin{eqnarray}
 q_0 &\approx &\frac{D}{2\gamma_F} \left( 1+
\sqrt{1-\left(\frac{2\gamma_F\kappa}{D}\right)^2}\, \right)\nonumber \\
&= &0.13 \AA^{-1}=0.07 {\rm rlu}. 
\end{eqnarray}
For $q<q_0$, $\Gamma_{\bf q} <\hbar \omega_{\bf q}$. 
A constant-{\bf q} scan yields a peak at a finite
energy (c.f., Fig.~\ref{fig_fwd}), demonstrating that
damped but still propagating spin waves exist.
For $q>q_0$, $\Gamma_{\bf q} >\hbar \omega_{\bf q}$ and the
spin excitation spectrum becomes overdamped 
(c.f., insert in Fig.~\ref{fig_gamma}). 
The spectrum in this part of
phase space is reminiscent of the paramagnon spectrum of 
conventional ferromagnetic 
materials at $T>T_C$\cite{MnSi_ishc}.
Specifically, while a constant-{\bf q} scan cutting through
$S({\bf q},\omega)$ yields no peak at a finite
energy, a constant-$\hbar\omega$ scan shows a peak at finite {\bf q}.
Some examples of such constant-$\hbar\omega$ scans for Fe$_{0.7}$Al$_{0.3}$
are shown in Fig.\ref{fig_cq}. It is interesting to note
that a crossover at finite {\bf Q}
from propagating to diffusive vibrational modes exists in
structural glasses\cite{gls_si}.

One source of damping is interactions between spin waves\cite{marsh}.
However, at $T=295$~K~$\approx 0.6T_C$, they would not {\em over}damp
spin waves at a very small wave vector 
$q\approx (0.07,0.07,0.07) =0.14q_{B.Z.}$, 
where $q_{B.Z.}=(\slantfrac{1}{2},\slantfrac{1}{2},\slantfrac{1}{2})$ 
is the Brillouin zone boundary. Furthermore,
damping due to spin wave interactions approaches zero when $T\rightarrow 0$,
while in Fe$_{0.7}$Al$_{0.3}$ damping increases upon cooling\cite{rsg_motoya}.
Therefore, spin wave interactions are not the main cause of damping
in Fe$_{0.7}$Al$_{0.3}$.
Strong damping in this material
is most likely due to
disorder introduced by random mixture of Fe and Al at the
$\gamma$ and $\beta$ sites, which at T$_f$ also freezes
the spins.

Finally, we compare magnetic excitations at 294~K in the 
FM phase to
those at 18~K in the SG phase. To facilitate such a comparison,
data are presented as $\chi''({\bf q},\omega)$ in Fig.~\ref{fig_cq}
to remove the thermal occupation factor [Eq.~(\ref{eq_fdt})].
Spin waves at small {\bf Q} near ${\bbox \tau}=(000)$ are known to 
become overdamped at low temperatures from previous cold neutron 
studies\cite{rsg_sfpp,rsg_murani,rsg_motoya,rsg_asbc,rsg_lms}.
This fact remains true near the Bragg point (111)\cite{note} and
is reflected in the enhanced intensity in the SG state (refer to the
open circles) for the $\hbar\omega=3$~meV
scans in Fig.~\ref{fig_cq}. The prevailing view about the diffuse
magnetic fluctuations is that frozen spins undergo uncorrelated
relaxations. However, unexpectedly, the strong {\bf q}-dependent
structure of $\chi''({\bf q},\omega)$ in the FM state 
remains in the SG state. In fact,
the $\chi''({\bf q},\omega)$ is identical at 18~K and at 
295~K for $\hbar\omega \ge 5$~meV. It appears that changes
in the dynamic magnetic spectrum in Fig.~(\ref{fig_map})
upon reaching the SG state occur mainly as a result of 
$q_0$ decreasing with temperature, so that when T approaches
zero, spectral weight at small {\bf q} and $\omega$ builds up
and no propagating features are present. The paramagnon-like
spectrum now covers the entire {\bf q}-$\omega$ space, indicating
remnant short-range spin clusters. It is interesting to note that 
the SG state with frozen disorder exhibits similar spin dynamics
as the conventional paramagnetic state where dynamic
disorder is caused by thermal energy.

In conclusion, we have characterized over a 
large $\omega$ and ${\bf q}$ range the  
anomalous spin dynamics in the FM phase of the
reentrant spin glass material Fe$_{0.7}$Al$_{0.3}$. Like the
SG state, the FM state seems also to be a mixed
state. In this case, ferromagnetic order and paramagnetic-like disorder
coexist at different length scales. In the SG state, the paramagnon-like
spin excitations dominate spin dynamics, suggesting short-range
spin clusters. This picture, from our experiment, now indicates that
the so-called FM state can possess a larger entropy.
There likely exists in the phase diagram a crossover between the
anomalous ferromagnetic state around 70\% Fe and the conventional 
ferromagnetic state at larger Fe concentrations.

We thank P. B\"{o}ni, G. Shirane, R. A. Cowley, J. D. Axe, J. Tranquada,
and A. Zheludev for useful discussions;
J. W. Lynn, S.-H. Lee and I. Zaliznyak for hospitality  
at NIST.
The work at BNL was supported by DOE under Contract 
No.\ DE-AC02-98CH10886 and in part by US-Japan program on Neutron
Scattering,
at LANL under the auspices of DOE.

\end{document}